\documentclass[12pt,twocolumn]{elsart}

\usepackage{graphicx}
\usepackage{times}

\begin{document}

\begin{frontmatter}
\title{Identification of High $\rm p_{\perp}$ Particles with the STAR-RICH Detector}

\author{The STAR-RICH Collaboration:}
\author[CERN]{A.~Braem},
\author[Bari]{D.~Cozza},
\author[CERN]{M.~Davenport},
\author[Bari]{G.~De~Cataldo},
\author[Bari]{L.~Dell~Olio},
\author[Bari]{D.~DiBari},
\author[CERN]{A.~DiMauro},
\author[YaRI]{J.~C.~Dunlop},
\author[YaHE]{E.~Finch},
\author[CERN]{D.~Fraissard},
\author[Bari]{A.~Franco},
\author[YaRI]{J.~Gans},
\author[Bari]{B.~Ghidini},
\author[YaRI]{J.~W.~Harris},
\author[YaRI]{M.~Horsley},
\author[YaRI]{G.~J.~Kunde},
\author[YaRI]{B.~Lasiuk},
\author[CERN]{Y.~Lesenechal},
\author[YaHE]{R.~D.~Majka},
\author[CERN]{P.~Martinengo},
\author[CERN]{A.~Morsch},
\author[Bari]{E.~Nappi},
\author[CERN]{G.~Paic},
\author[CERN]{F.~Piuz},
\author[Bari]{F.~Posa},
\author[CERN]{J.~Raynaud},
\author[YaRI]{S.~Salur},
\author[YaHE]{J.~Sandweiss},
\author[CERN]{J.~C.~Santiard},
\author[YaRI]{J.~Satinover},
\author[CERN]{E.~Schyns},
\author[YaRI]{N.~Smirnov},
\author[CERN]{J.~Van~Beelen},
\author[CERN]{T.~D.~Williams},
\author[YaHE]{Z.~Xu}
% D.~Cozza$^{1}$, M.~Davenport$^{2}$, G.~De~Cataldo$^{1}$, L.~Dell~Olio$^{1}$,
% D.~DiBari$^{1}$, A.~DiMauro$^{2}$, J.~C.~Dunlop$^{3}$, E.~Finch$^{3}$,
% D.~Fraissard$^{2}$, A.~Franco$^{1}$, J.~Gans$^{3}$, B.~Ghidini$^{1}$,
% J.~W.~Harris$^{3}$, M.~Horsley$^{3}$, G.~J.~Kunde$^{3}$,
% B.~Lasiuk$^{3}$, Y.~Lesenechal$^{2}$, R.~D.~Majka$^{3}$, P.~Martinengo$^{2}$,
% A.~Morsch$^{2}$, E.~Nappi$^{1}$, G.~Paic$^{2}$, F.~Piuz$^{2}$, F.~Posa$^{1}$,
% J.~Raynaud$^{2}$, S.~Salur$^{3}$, J.~Sandweiss$^{3}$, J.~C.~Santiard$^{2}$,
% J.~Satinover$^{3}$, E.~Schyns$^{2}$, N.~Smirnov$^{3}$, J.~Van~Beelen$^{2}$,
% T.~D.~Williams$^{2}$, Z.~Xu$^{3}$}

\address[CERN]{CERN HMPID group, CERN, Geneva CH-1211}
\address[Bari]{Bari HMPID Group, Bari, Sez. INFN and Dipartimento di fisica, 70124, Italy}
\address[YaRI]{Yale RHI Group, New Haven, CT 06520-8124, USA}
\address[YaHE]{Yale HE Group, New Haven, CT 06520, USA}

\begin{abstract} 
 The STAR-RICH detector
 extends the particle identification
 capabilities of the STAR experiment for charged hadrons 
 at mid-rapidity.  This detector represents the first use of 
 a proximity-focusing CsI-based RICH detector in a collider experiment.
 It provides identification
 of pions and kaons up to 3~GeV/c and protons up to
 5~GeV/c.  The characteristics and performance of the device
 in the inaugural RHIC run are described.
\end{abstract}
\end{frontmatter}

\section{Introduction}
Penetrating probes are of importance for the study of the dense
partonic matter formed after the collision of heavy nuclei at RHIC
energies.  At RHIC a new regime is reached, in which the collisions are
semihard, producing so-called minijets.  In nucleon-nucleon collisions
the behavior of minijets and the spectra of the particles emitted in
minijets are amenable to perturbative QCD calculations, and thus
represent a firm ground for comparison with experiments.  In nucleus-%
nucleus collisions particles produced in the initial state of the
collision (before thermalization) are used to ``probe'' the matter
formed in the collision.  By measuring the momentum spectra of
identified particles a picture characteristic of the matter they have
traversed will be obtained~\cite{dedx}.  This makes, among other
probes, the measurement of identified particles at high momentum an
essential objective of heavy-ion detectors.

To achieve this goal the STAR detector has complemented its particle
identification at low momenta (achieved with the TPC and SVT until 0.6
and 1~GeV/c, for kaons and protons, respectively) with a patch of
detector dedicated to high momentum identification, based on the
identification of photon patterns emitted by Cherenkov radiation.

The STAR-RICH proximity-focusing Ring Imaging Cherenkov (RICH)
detector was originally built as a 1~$\rm m^2$ prototype for the HMPID
(High Momentum Particle Identification) detector of the ALICE
experiment at the LHC~\cite{piuz1}.  After extensive and fully
satisfactory tests at the H4 line of the CERN SPS it was decided to
install it in STAR~\cite{starrich:proposal}.  The STAR-RICH detector
covers 2\% of the TPC acceptance and has been installed in the central
rapidity region covering $\Delta \eta < 0.3$, for event vertices at
the center of the TPC, and $\Delta \phi = 20{\rm {}^o}$.  Using this
detector, the momentum range of particle identification should be
extended to 3~GeV/c for kaons and 5~GeV/c for protons.  The present
detector represents the first use of a proximity-focusing RICH
detector, with a MWPC pad cathode coated with CsI, in a collider
experiment.  In the following we will describe the detector and its
performance.  In the last part we will present the pattern recognition
algorithm applied to the data.

%%%%%%%%%%%%%%%%%%%%%%%%%%%%%%%%%%%%%%%%%%%%%%%%%%%%%%%%%%%%%%%%%%%%%%%%%%%%
\begin{figure}[htb]
\begin{center}
\includegraphics[width=.45\textwidth]{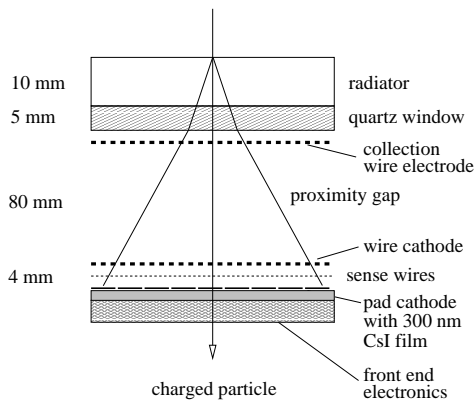}
\caption{Cutaway view of the detector\cite{piuz1}.}
\label{fig:detector}
\end{center}
\end{figure}
%%%%%%%%%%%%%%%%%%%%%%%%%%%%%%%%%%%%%%%%%%%%%%%%%%%%%%%%%%%%%%%%%%%%%%%%%%%%

%%%%%%%%%%%%%%%%%%%%%%%%%%%%%%%%%%%%%%%%%%%%%%%%%%%%%%%%%%%%%%%%%%%%%%%%%%
\section{Characteristics}
A schematic of the detector is shown in figure~\ref{fig:detector}.  
The Cherenkov photon radiating medium is a 1~cm thick layer of liquid
perfluorohexane ($\rm C_6 F_{14}$) circulated into a tray, closed by a
0.5~cm thick UV-grade fused silica plate in order to match the
spectral range of the photoconverter.  Two such trays of
1330x413~mm${}^2$ size are implemented in this detector.

Using a proximity RICH geometry, the distance radiator to
photodetector is fixed at 80~mm for optimization of pattern
recognition~\cite{piuz3}.  Electrons released by ionizing particles in
the proximity gap are prevented from entering the photodetector volume
by a positive polarization of the collection electrode close to the
radiator.

The photodetector is made of a conventional MWPC having one of the
cathode plane segmented into pads of 8.0x8.4~mm${}^2$ area at a
distance of 2~mm from the anode plane, which consists of 20~$\rm \mu
m$ diameter wires spaced by 4.2~mm.  The opposite cathode consists of
100 $\rm \mu m$ wires, spaced by 2.1~mm and located at 2.2~mm from the
anode plane.  Four independent panels, of 640x400~mm${}^2$ sensitive
area, form the cathode plane, resulting in a total active area of
1280x800~mm${}^2$.

The photoconverter consists of a thin layer of CsI deposited under
vacuum on the pad cathode surface.  The useful spectral range of such a
photocathode is limited between the CsI threshold at 205~nm and the
radiator transparency at 170~nm.  It has been demonstrated that such a
photocathode can be operated under gas at atmospheric pressure with a
yield of 17$\pm$2 reconstructed photoelectron clusters per ring for
$\beta\sim 1$, at a gas gain of
$0.8-2.\times10^{5}$~\cite{piuz1,piuz3,piuz2,an96}.  If kept under
inert gas flow, the photocathode has a quantum efficiency stable over
several years.  It can be operated in an ``open MWPC geometry''
without suffering from photon feedback.  An accurate localization of
the particles can be achieved by charge centroid calculation with this
photocathode.

The electronics readout is implemented at the back of the pad panels.  
Given the modest event rate expected around ion colliders, it is based
on the ASIC GASSIPLEX chip providing 16 analog multiplexed
channels~\cite{gassiplex}.  In the STAR-RICH application 60 chips (960
channels) are daisy chained in order to minimize the number of
expensive ADCs while keeping the readout time compatible with the RHIC
interaction rate.  The digitization and zero-suppression are performed
by using the commercial CAEN V550 ADC VME module.  The multiplexing
frequency is 1~MHz and the noise level on the detector is of the order
of 1000 e- r.m.s over 15360 channels in total.

A detailed description of these elements and of the beam test
measurements performed can be found in
references~\cite{piuz1,piuz3,piuz2,an96,HMPID:TDR}.

%%%%%%%%%%%%%%%%%%%%%%%%%%%%%%%%%%%%%%%%%%%%%%%%%%%%%%%%%%%%%%%%%%%%%%%%%%%%
\begin{figure}[htb]
\begin{center}
\includegraphics[width=.4\textwidth]{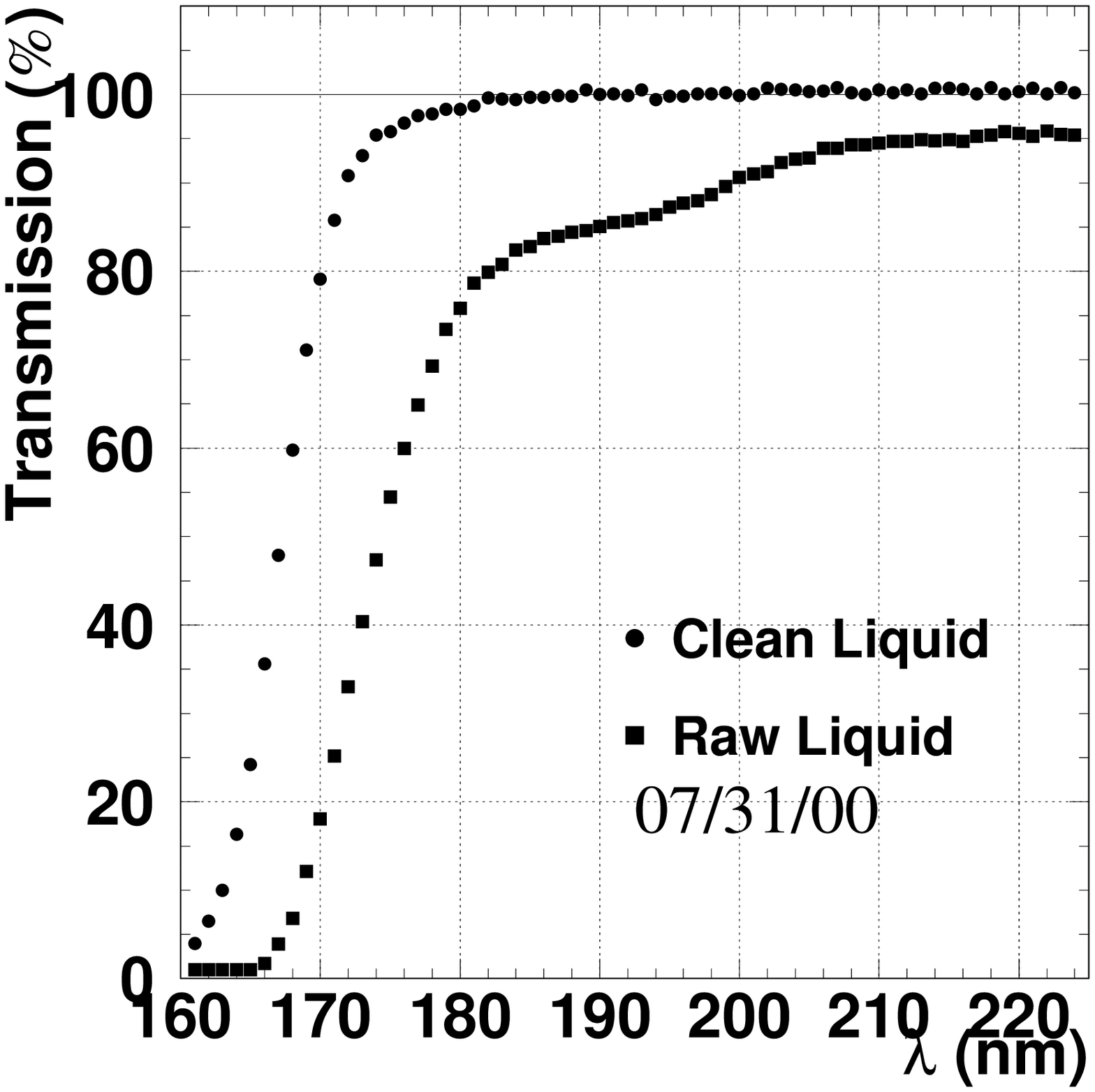}
\includegraphics[width=.4\textwidth]{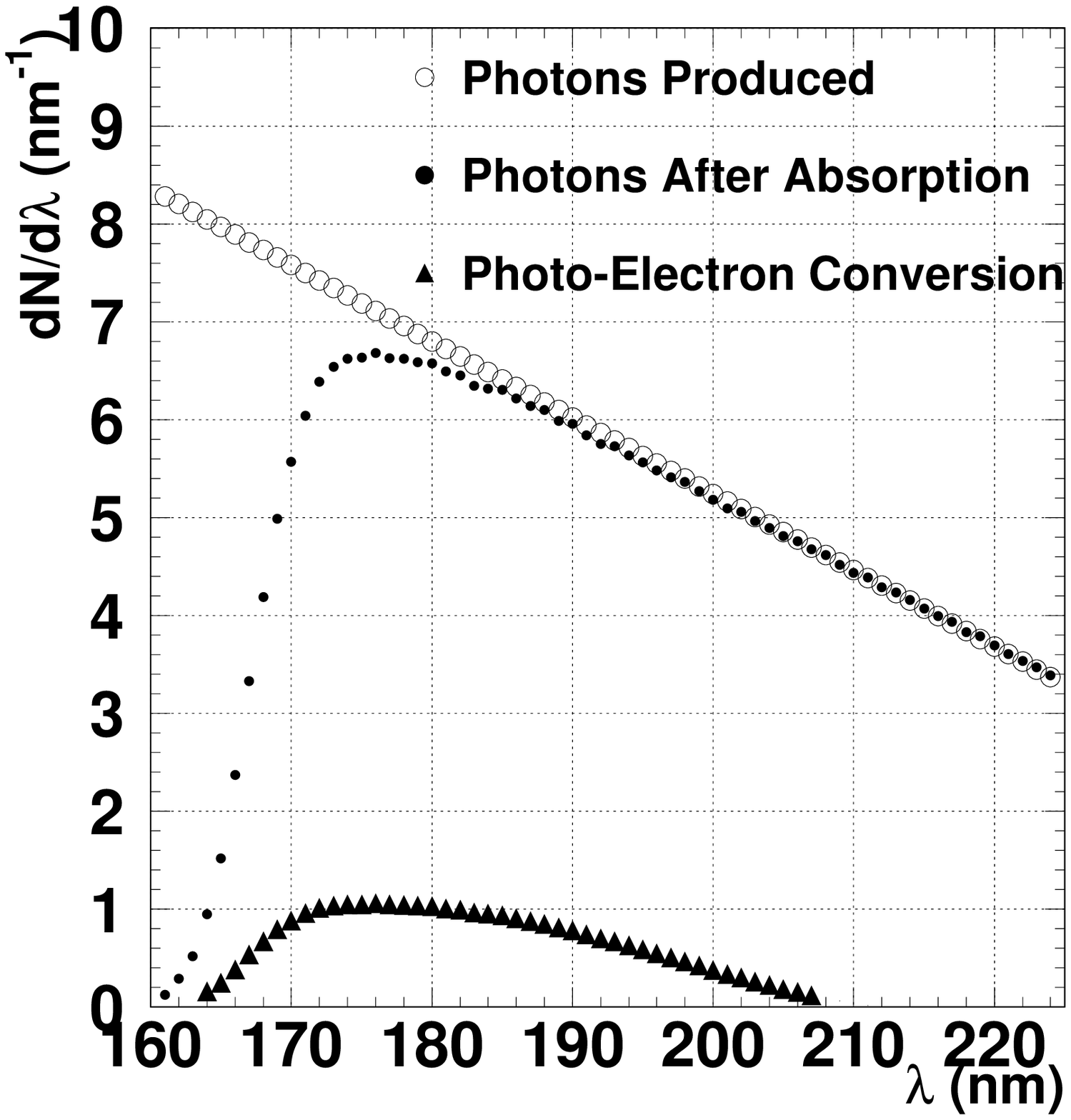}
\caption{\footnotesize Top: the measured
 transmission of the liquid
 radiator in raw (squares) and cleaned (circles) form.
 Bottom: the simulated spectrum of Cherenkov
 light produced in the
 liquid radiator (open circles) and the effect of the finite
 liquid and quartz transmission.  The resulting spectrum
 transmitted to the pad plane (solid circles) is folded
 with the quantum efficiency of the CsI to show the
 spectrum of photons converted to electrons (triangles).}
\label{fig:photons}
\end{center}
\end{figure}
%%%%%%%%%%%%%%%%%%%%%%%%%%%%%%%%%%%%%%%%%%%%%%%%%%%%%%%%%%%%%%%%%%%%%%%%%%%%

CsI has a photo-electric threshold of 205~nm (6.0 eV) which restricts
its sensitivity to vacuum ultra-violet (VUV) photons.  This imposes
constraints on the spectral properties of the radiator medium, its
containment vessel, as well as the MWPC gas.  In order to extend the
PID capabilities of the TPC, a radiator with an index of refraction in
the range of 1.2-1.3 is necessary.  Perfluorohexane (C$_{6}$F$_{14}$)
has an index of refraction of 1.3 at 170~nm~\cite{HMPID:TDR}, and
shows transparency to VUV radiation down to 152~nm, if water and
oxygen are reduced to the ppm level.  Pure methane is utilized as the
MWPC gas since it is transparent to VUV radiation down to 130~nm and
allows the chamber to be run in a stable manner at a gas gain of $\sim
10^{5}$ with no deterioration of quantum efficiency up to a rate
density of $\rm 2.0\times 10^{4} s^{-1} cm^{-1}$~\cite{an96}.  The
importance of a transparent radiator medium is depicted in
figure~\ref{fig:photons}.  The top panel shows the measured
transmission of the liquid before and after cleaning while the bottom
panel shows the simulated spectrum of Cherenkov light incident on the
pad plane and subsequently converted to photo-electrons.

In addition to the spectral constraints CsI imposes, it is also
hygroscopic and care must be taken to isolate it from water, as well
as oxygen.  From experience accumulated over the past 5 years and six
large CsI photocathodes, the following observations can be reported.  
Keeping the CsI photocathode under permanent flow of Argon (10-15~l/h)
in a protective sealed metallic enclosure of volume 1.5 l with
moisture levels of less than 20 ppm resulted in a reduction of less
than 10\% in quantum efficiency over 5 years.  Exposing a CsI
photocathode to 200~ppm moisture for 30 hours without argon flow and
to $1.10\times10^5$~ppm oxygen for 20 hours had no effect on the
quantum efficiency, following tests done at CERN test beams with the
quantum efficiency measured before and after exposure to contaminants.
This last point was a significant consideration in transporting the
detector from CERN to BNL, providing constraints on the construction
of a sealed transportation vessel~\cite{emile}.

%%%%%%%%%%%%%%%%%%%%%%%%%%%%%%%%%%%%%%%%%%%%%%%%%%%%%%%%%%%%%%%%%%%%%%%%%
\section{Performance and Analysis}
The STAR-RICH capability to identify charged particles has been
studied in central Au+Au collisions ($\sqrt{s_{NN}}$ = 130 GeV) during
the first year of running of RHIC, where about 900k central and 900k
minimum bias triggers have been recorded.  The identification of high
momentum charged particles in STAR (p$>$1~GeV/c) can be performed once
the momentum and the velocity, $\beta$, for each particle, are
measured.  The STAR-RICH detector allows for measurement of the
$\beta$ of the particle by the reconstruction of the Cherenkov angle,
while the momentum of the track is obtained from the TPC.
%%%%%%%%%%%%%%%%%%%%%%%%%%%%%%%%%%%%%%%%%%%%%%%%%%%%%%%%%%%%%%%%%%%%%%%%%%%%
\begin{figure}[htb]
\begin{center}
\includegraphics[width=0.4\textwidth]{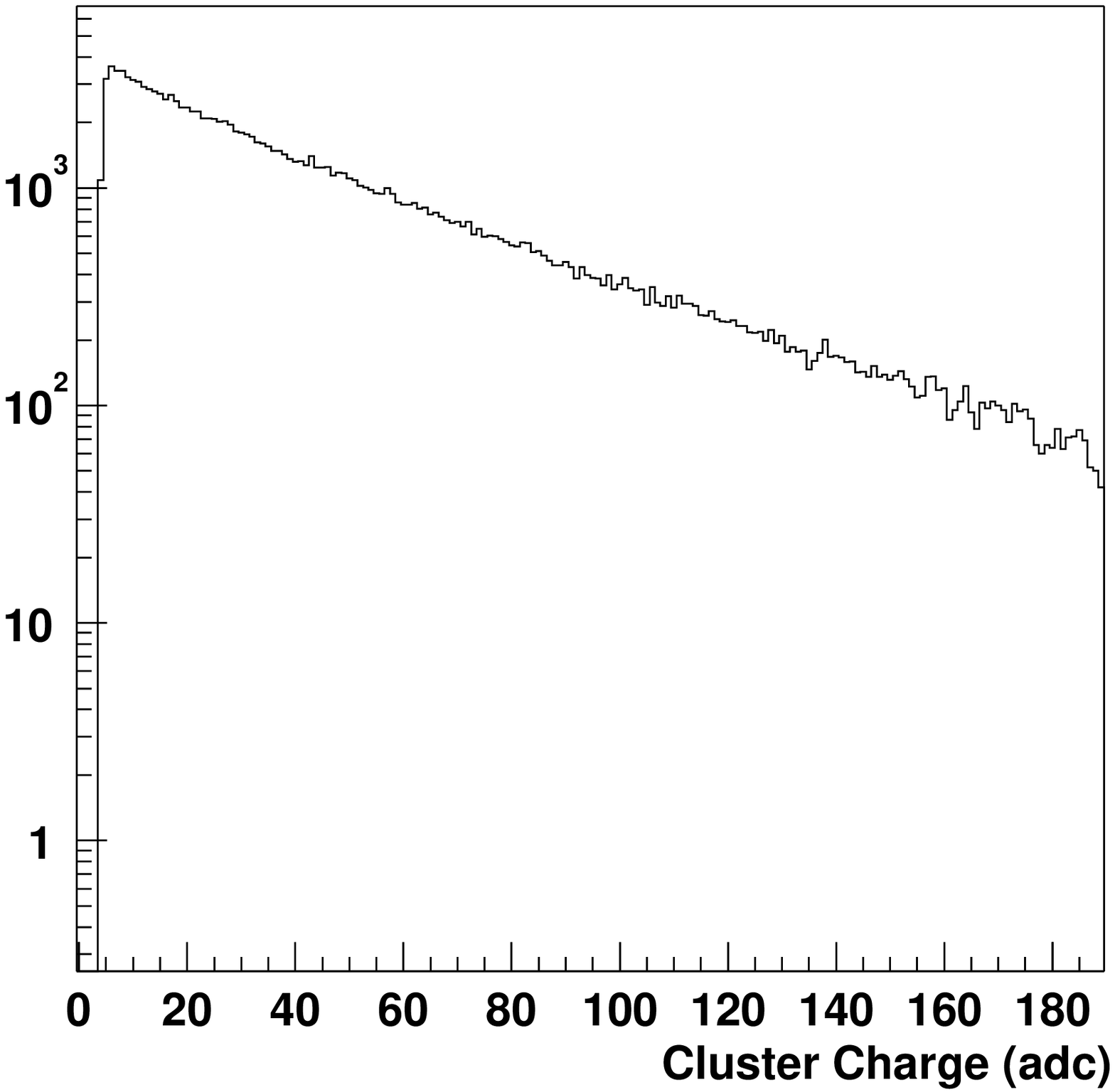}
\includegraphics[width=0.4\textwidth]{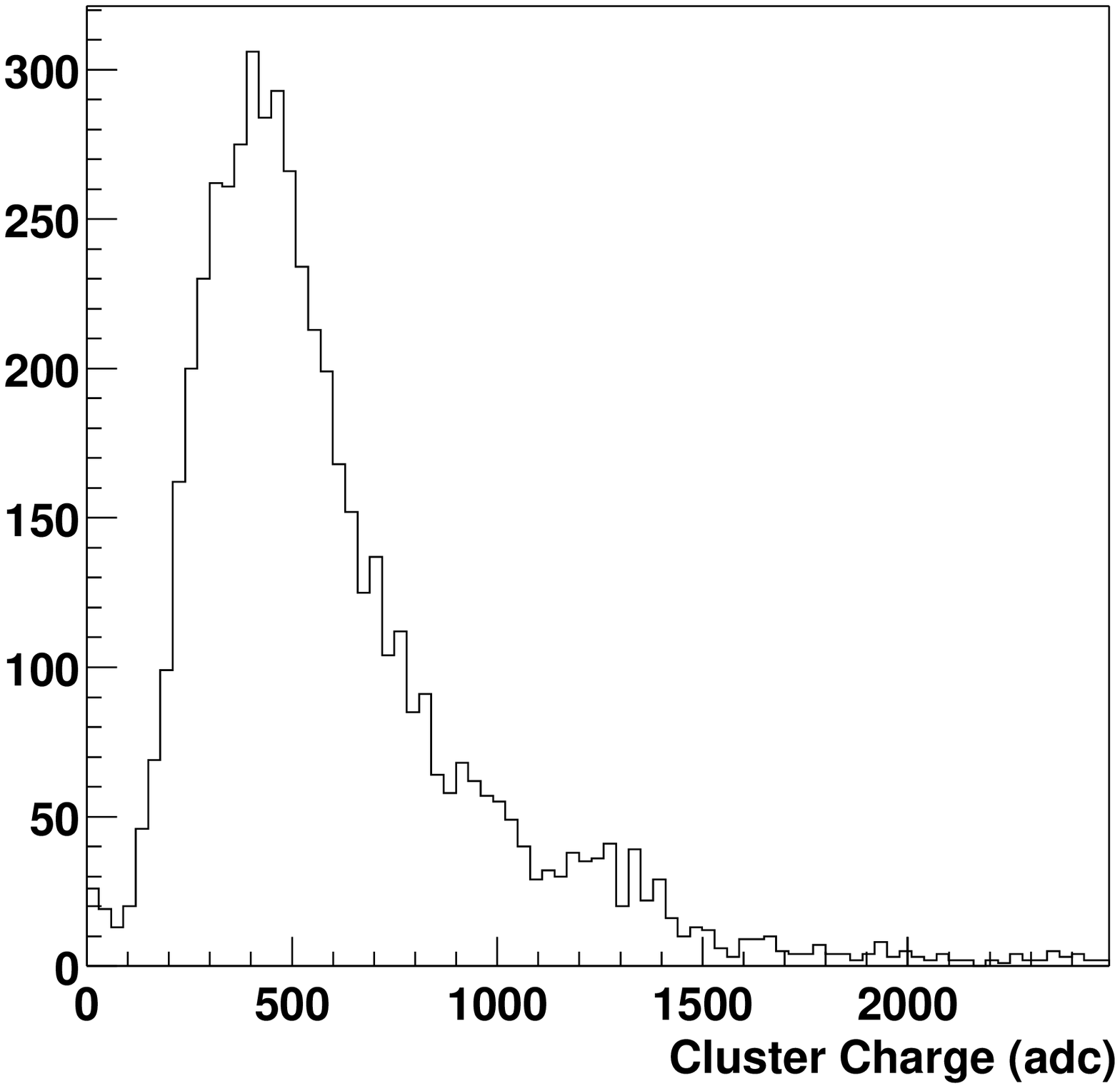}
\caption{
 Charge spectrum of reconstructed clusters in the chamber.  Top: hits
 within the expected fiducial area for photons.  Bottom: hits within
 0.5 cm of a projected TPC track.  Overflows affect the spectrum
 above $\sim$950 ADC counts.  }
\label{fig:hits}
\end{center}
\end{figure}
%%%%%%%%%%%%%%%%%%%%%%%%%%%%%%%%%%%%%%%%%%%%%%%%%%%%%%%%%%%%%%%%%%%%%%%%%%%

Extrapolation of tracks to the RICH show a distribution of residuals,
between the projected impact point of the track and the centroid of a
cluster with charge Q$>$ 120 ADC (see figure~\ref{fig:hits}), with a
$\sigma$ of 4~mm for tracks with a momentum greater than 1~GeV/c.  
Tracks with smaller momentum have a much larger width since multiple
scattering becomes more appreciable.  Tracks with residuals $<$ 5 mm
are selected for further analysis.  This requirement mainly rejects
those particles that, although reconstructed in the TPC, are not
detected in the STAR-RICH detector because they either undergo an
upstream interaction with the material between the TPC and the
STAR-RICH or decay.

%%%%%%%%%%%%%%%%%%%%%%%%%%%%%%%%%%%%%%%%%%%%%%%%%%%%%%%%%%%%%%%%%%%%%%%%%%%%
\begin{figure}[tb]
\begin{center}
\includegraphics[height=.45\textwidth,angle=270]{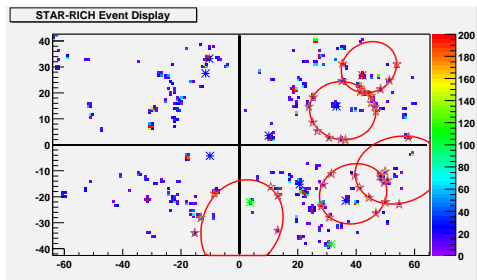}
\caption{
 \footnotesize An off-line event display of the RICH detector.  
 Colored squares show the ADC value for pads with charge above
 threshold.  Reconstructed rings are shown for associated tracks.}
\label{fig:event}
\end{center}
\end{figure}
%%%%%%%%%%%%%%%%%%%%%%%%%%%%%%%%%%%%%%%%%%%%%%%%%%%%%%%%%%%%%%%%%%%%%%%%%%%
After having selected a track candidate, an algorithm for the
recognition of Cherenkov patterns based on the Hough transform method
is applied.  It consists of a mapping of the pad coordinate space
directly to the parameter space of the Cherenkov photon angle,
extracted from the photon cluster coordinate by a geometrical
back-tracking.  More details can be found in reference~\cite{nico}.  
This method is efficient even in the presence of large occupancy.
In central collisions in STAR, however, the pad occupancy
never exceeds 5\%, corresponding to a good Cherenkov photon signal to
noise ratio.  The pattern generated by the detected photoelectrons is
a function of the track incident angle.  A mean value of about 7$\rm
{}^o$ is obtained for tracks with $\left|\eta\right| <$ 0.15 coming
from a primary vertex with $\rm \left|vertex~z\right| <$ 50 cm.  This
produces Cherenkov patterns with elliptical shapes.  
Figure~\ref{fig:event} shows the offline display of an event in the
RICH detector, where the reconstructed rings associated to tracks with
momentum $>$ 1~GeV/c have been also drawn.

\begin{figure}[htb]
\begin{center}
\includegraphics[height=.45\textwidth,angle=270]{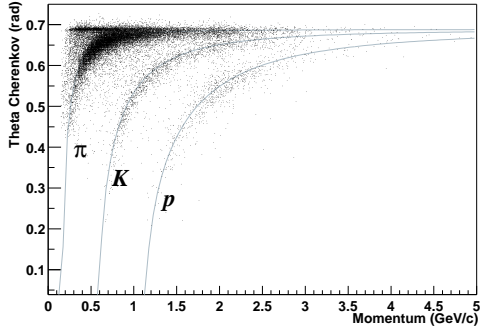}
\caption{Reconstructed Cherenkov angle in the RICH vs momentum from the TPC.}
\label{fig:changle}
\end{center}
\end{figure}

\begin{figure}[htb]
\begin{center}
\includegraphics[height=.45\textwidth,angle=270]{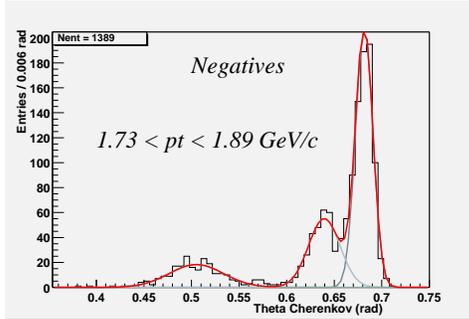}
\caption{
 Reconstructed Cherenkov angle in the RICH in a restricted range of
 transverse momentum measured by the TPC.}
\label{fig:tightchangle}
\end{center}
\end{figure}

The reconstructed Cherenkov angle as a function of the track momentum
is shown in figure~\ref{fig:changle}.  Bands with low background,
centered around the predicted curves for pions, kaons, and protons,
can clearly be seen.  Figure~\ref{fig:tightchangle} shows an example
of the distribution of the reconstructed Cherenkov angle, in a small
interval of transverse momentum, where the three peaks corresponding
to pions, kaons, and protons are clearly visible.  This distribution
is fitted in order to extract the relative yields.
%%%%%%%%%%%%%%%%%%%%%%%%%%%%%%%%%%%%%%%%%%%%%%%%%%%%%%%%%%%%%%%%%%%%%%%%%%%%
%%%%%%%%%%%%%%%%%%%%%%%%%%%%%%%%%%%%%%%%%%%%%%%%%%%%%%%%%%%%%%%%%%%%%%%%%%%

In an alternative method, three mass hypotheses are made for a given
track ($\pi$, K, and p).  The limiting bounds of the Cherenkov rings,
defined by the dimensions of the radiator vessel, are projected on to
the pad plane.  The mass of the particle can be assigned by selecting
the band with the highest density of photons, from which yields can be
extracted.
\section{Conclusion}
A proximity-focusing CsI RICH detector, developed by the ALICE
collaboration, has been installed in the STAR detector at RHIC,
representing the first use of such a device in a collider experiment.  
This device extends the PID capabilities of the spectrometer to
3~GeV/c for pions and kaons, and 5~GeV/c for protons.  The performance
of the detector is within expectations, which will allow an
examination of the effect of matter at high density on high $\rm
p_{\perp}$ particles in the coming years.

% This detector has produced an anti-proton to proton ratio
%in the transverse momentum range 2.0$<$p$_T<$2.5 (GeV/c) at mid-rapidity
%for 130~GeV/A Au-Au collisions.  This measurement will be extended with
%further data in the coming year which will allow an examination of
%the effect of high density matter on the fragmentation distribution
%of high p$_T$ partons.

%%%%%%%%%%%%%%%%%%%%%%%%%%%%%%%%%%%%%%%%%%%%%%%%%%%%%%%%%%%%%%%%%%%%%%%

\end{document}